\documentclass[useAMS,usenatbib]{mn2e}

\usepackage{color}
\usepackage{lscape,graphicx}
\usepackage{amsmath}
\usepackage{journal_shortcuts}
\usepackage{amssymb}
\usepackage{multirow}

\def\ga{{\rm\thinspace gauss}}

\def\j1431{\hbox{MaxBCG J217.95869+13.53470}}

\def\h0{\hbox{{\rm H}$^0$}}
\DeclareMathAlphabet{\vib}{OML}{cmm}{m}{it}
\newcommand*{\satellite}[1]{\textit{#1}}
\newcommand*{\xmm}{\satellite{XMM-Newton}}

\newcommand{\lsim}{\mathrel{\hbox{\rlap{\lower.55ex\hbox{$\sim$}} \kern-.3em \raise.4ex \hbox{$<$}}}}
\newcommand{\gsim}{\mathrel{\hbox{\rlap{\lower.55ex\hbox{$\sim$}} \kern-.3em \raise.4ex \hbox{$>$}}}}

\title[X-ray evidence of a shock at the Coma relic]{First X-ray evidence for a shock at the Coma relic}
\author[G.~A.~Ogrean et al.]{G.~A.~Ogrean$^{1}$\thanks{E-mail: gogrean@hs.uni-hamburg.de}, M. Br{\"u}ggen$^{1}$\\
$^{1}$Hamburger Sternwarte, Gojenbergsweg 112, 21029, Hamburg, Germany}
\begin{document}

\date{Accepted xxxx xx xx. Received xxxx xx xx; in original form xxxx xx xx}

\pagerange{\pageref{firstpage}--\pageref{lastpage}} \pubyear{2012}

\maketitle

\label{firstpage}

\begin{abstract}

The Coma cluster is one of the nearest galaxy clusters, and the first one in which a radio halo and a peripheral relic were discovered. While its halo and the central parts of the intracluster medium have been studied extensively, X-ray observations of the plasma near its relic have been scarce. Here, we present results from a re-analysis of a 22-ks archival \xmm\ observation. Across the relic, we detect a shock of Mach number $\sim 2$. This excludes the previously suggested hypothesis that the relic was formed by turbulence. Furthermore, multiwavelenth observations and numerical models do not support the scenario in which the shock at the Coma relic is an outgoing cluster-merger shock. Instead, our results lend support to the idea that the relic coincides with an infall shock front formed just as the NGC 4839 group falls onto the cluster along a cosmic filament.

\end{abstract}


\section{Introduction}

Over the last 40 years, diffuse, Mpc-scaled, steep-spectrum radio emission has been detected in the outskirts of a number of galaxy clusters \citep[e.g., ][]{Ferrari2008,vanWeeren2009,vanWeeren2011}. These radio objects, named radio relics, have an irregular, arc-shaped morphology, generally show a spectral index gradient across their surface, and are believed to be formed via acceleration at merger shock waves.

The Coma cluster is one of the nearest galaxy clusters ($z=0.0231$), and the first one in which a peripheral radio relic was discovered \citep{Ballarati1981}. However, X-ray observations of the plasma near the relic have been scarce. The Coma relic, radio source $1253+275$, has a length of roughly 2 Mpc \citep{BrownRudnick2011}, and is located SW of the cluster, at a projected radius of approximately $0.8r_{\rm 200}$. Using galaxy optical catalogues and 21-cm line observations taken with the 305-m radio telescope of the Arecibo observatory, \citet{Fontanelli1984} discovered that the galaxy distribution in the region of the Coma cluster delineates a NE-SW-oriented cosmic filament, which includes the Coma cluster itself. In the context of matter accreting along this filament onto the Coma cluster, and given our current understanding of radio relic formation, the presence of a radio relic at the SW edge of Coma is not surprising. Yet, the formation mechanism of the relic remains controversial \citep{Ensslin1998,FerettiNeumann2006,BrownRudnick2011}, and understanding it requires multiwavelength, high-quality observations. 

There has been only one attempt of detecting a shock at the Coma relic, using an \xmm\ observation carried out in 2003 \citep{FerettiNeumann2006}. The analysis of the \xmm\ data revealed a constant gas temperature of $\sim 3$ keV at and on the inner side of the relic. If the Coma relic can be explained by a shock front, then the temperature would be expected to increase at the relic, and then decrease rapidly by a factor of approximately $(5\mathcal{M}^2+14)/16$, where $\mathcal{M}$ is the shock Mach number \citep[e.g.,][]{PaulIapichino2011}. In the absence of a temperature increase or jump, \citet{FerettiNeumann2006} suggested that the relic was instead formed by turbulence triggered as matter accretes onto Coma. 

At the relic, the main source of X-ray emission is not the Coma cluster itself, but a smaller group of galaxies located along the NE-SW filament, NE of the relic. By comparing optical and X-ray observations at the relic, \citet{Neumann2001} have shown that the galaxies belonging to the group, including its brightest central galaxy, NGC 4839, are on average situated approximately 300 kpc closer to the Coma centre than the group's X-ray halo. This displacement suggests that the subgroup is falling into the Coma cluster \citet{Neumann2001}.

More recently, \citet{BrownRudnick2011} studied the galaxy surface density and redshift distribution around the relic, using Sloan Digital Sky Survey (SDSS) data. They identified a 2-Mpc-long ``wall'' of galaxies on the inner side of the relic, infalling towards Coma's centre with a velocity of approximately 500 ${\rm km\,s^{-1}}$. The length and position angle of the ``wall'' are very similar to those of the relic. They argue that the Coma relic traces a stationary infall shock formed behind the infalling group of galaxies. Infall shocks can form only in the early stages of a merger event, unlike (also stationary) accretion shocks, which occur due to continuous accretion onto the cluster.

Here, we present results from a re-analysis of the \xmm\ observation at the relic, which uses a more careful background treatment then the one of \citet{FerettiNeumann2006}. Throughout the Letter, we assume a flat $\Lambda$CDM cosmology, with $H_{\rm 0} = 70$ km s$^{-1}$ Mpc$^{-1}$, $\Omega_{\rm M}=0.3$, and $\Omega_{\rm \Lambda} = 0.7$. As X-ray emission at the relic comes mostly from the infalling subgroup, our analysis assumes a gas redshift of $0.0246$, equal to the redshift of NGC 4839 \citep{Smith2000}, which is the brightest cluster galaxy in the infalling group. At this distance, 1 arcmin translates to $29.7$ kpc. All errors are $1\sigma$ statistical errors, unless otherwise noted.

\section{Observations and data reduction}
\label{s:observations}

\begin{table*}
  \caption{Level of residual SP contamination and ``clean'' exposure times. $\mathcal{R}$ describes the ratio of count rates inside and outside the FOV, in the hard-energy band. See text for details.}
  \begin{center}
    \begin{tabular}{ccccccc}
      \hline
	ObsID      & $\mathcal{R}$ MOS1 & $\mathcal{R}$ MOS2 & $\mathcal{R}$ pn & Exp. time MOS1 & Exp. time MOS2 & Exp. time pn \\
      \hline
	0058940701 & $1.272\pm 0.094$   & $1.046\pm 0.070$   & $1.207\pm 0.119$ & $16.0$         & $16.5$         & $9.2$ \\
	0406610301 & $1.074\pm 0.081$   & $1.167\pm 0.081$   & $1.210\pm 0.115$ & $10.3$         & $10.3$         & $6.3$ \\
      \hline
    \end{tabular}
  \end{center}
  \label{tab:ressp}
\end{table*}

This Letter is based on a 22-ks archival \xmm\ observation (ObsID 0058940701) centred on 1253+275 -- the SW relic in the outskirts of the Coma cluster. The observation was taken on June 10, 2003, using the thin filter. ICM emission from the Coma and the infalling subgroup is present across the whole field of view (FOV). Therefore, an additional observation was required to model the sky background. We used ObsID 0406610301, a 15-ks, thin-filter observation of the quasar HS 1251+2636, with an aim point angular distance of 0.92 degrees from the Coma relic, in the SW direction. Both datasets were reduced with the Extended Source Analysis Software ({\sc esas}) integrated in the Science Analysis System ({\sc sas}) version 12.0.1, and the most recent calibration files as of November 2012. The data processing steps closely follow those suggested in the \xmm\ {\sc esas} cookbook~\footnote{ftp://xmm.esac.esa.int/pub/xmm-esas/xmm-esas.pdf}.

Additionally, we used NASA's HEASARC X-ray Background Tool~\footnote{http://heasarc.gsfc.nasa.gov/cgi-bin/Tools/xraybg/xraybg.pl} to extract a \emph{ROSAT} All-Sky Survey (RASS) background spectrum from the same region as the spectrum of ObsID 0406610301. The RASS spectrum allows us to better constrain the LHB normalization. However, we note that excluding the RASS spectrum from the spectral fit does not siginificantly change any of the best-fit background parameters, with the exception of the LHB normalization; the LHB normalization contributes little to the fit though, given the very low LHB temperature.

We checked for residual soft proton (SP) contamination by comparing the hard-band ($10-12$ keV for MOS, $12-14$ keV for pn) count rates inside and outside the field-of-view (FOV). \citet{delucamolendi2004} found that count rate ratios (in/out; $\mathcal{R}$) below 1.15 indicate event files that are essentially not contaminated by residual soft protons, while ratios between 1.15 and 1.3 translate to slight SP contamination. The higher the ratio, the larger the residual SP contamination. In Table \ref{tab:ressp} we list $\mathcal{R}$ for the flare-filtered EPIC event files of the two observations, along with the ``clean'' exposure times.

The quiescent particle background (QPB) was modelled spatially and spectrally, and subtracted from all the images and spectra. The modelling was done in three steps: (1) the spectral parameters of the detector corners are determined; (2) filter-wheel closed (FWC) observations with similar spectral parameters are identified in the archived-observation database; (3) the FWC data is scaled by the ratio of the observation corner spectra to FWC corner spectra. Finally, {\sc esas} creates images and spectra of the QPB using the scaled FWC data.

\section{Analysis}
\label{s:analysis}

\subsection{The sky background}
\label{s:bkgmod}

Sky background spectra were extracted from the ObsID 0406610301 event files in circular annuli between 6 and 12 arcmin around the optical axis, and binned to a minimum of 30 counts per spectral channel. The spectra were modelled as the sum of Local Hot Bubble (LHB), Galactic Halo (GH), Cosmic X-ray Background (CXB), and residual SP (RESP) emission. The LHB and the GH were described by thermal components, while the CXB and the RESP were described as power-laws. For the absorbed components (GH, CXB), the X-ray column density was fixed to the weighted average Galactic H{\sc i} column density in the Leiden/Argentine/Bonn (LAB) Survey. Two constants were introduced to describe the solid angle corresponding to the extraction region, and to account for calibration errors between the EPIC detectors (calculated with respect to MOS2). The spectral model is similar to that of \citet{Snowden2008}, and described there in more detail. Table \ref{tab:bkgsp} summarizes the best-fit sky background model. Despite our attempt to model RESP emission, none was detected in any of the spectra; this is not necessarily surprising, given that, within the error bars, $\mathcal{R}$ of all detectors is consistent with $<1.15$. The fit was obtained with {\sc XSpec} v12.7.1 in the energy band $0.5-7$ keV (excluding the $1.2-1.9$ keV band, which is contaminated by fluorescent instrumental lines), and has $\chi^2/{\rm d.o.f.}=1.02$.

\begin{table}
 \caption{Best-fit background model. Temperatures are in units of keV. Normalizations are in default {\sc XSpec} units per square arcmin. Power-law normalizations are measured at 1 keV.}
\begin{center}
  \vspace{0.2cm}
 \begin{tabular}{lccc}
   \multicolumn{4}{l}{\footnotesize{\textsc{SPECTRAL MODEL:}}} \\ 
   \multicolumn{4}{l}{\footnotesize{\textsc{calib $\times$ solid\_angle $\times$ ( apec + ( apec + pow ) $\times$ phabs )}}} \\ 
  \hline
   Component & $\Gamma$ & T & Normalization \\
  \hline
    LHB & -- & $0.08^\dag$ & $(1.4\pm 0.23) \times 10^{-6}$ \\
    GH  & -- & $0.18_{-0.0054}^{+0.0057}$ & $8.9_{-0.81}^{+0.84} \times 10^{-7}$ \\
    CXB & $1.41^\dag$ & -- & $7.3_{-0.54}^{+0.56} \times 10^{-7}$ \\
  \hline\hline
	& \multicolumn{2}{c}{MOS1} & pn \\
  \hline
    {\sc calib} & \multicolumn{2}{c}{$1.25_{-0.10}^{+0.11}$} & $1.02_{-0.065}^{+0.072}$ \\
  \hline
  \multicolumn{4}{p{0.4\textwidth}}{$\dag$ \footnotesize{frozen}} 
 \end{tabular}
\end{center}
 \label{tab:bkgsp}
\end{table}

\subsection{Evidence of a shock}
\label{sec:icmmodelling}

To the best-fit background model above, we added an absorbed, single-temperature thermal component to model emission from the ICM. Additionally, we used a power-law to model RESP emission. The X-ray column density was fixed to $8.06\times 10^{19}$ atoms cm$^{-2}$ -- the weighted average LAB value measured in a circle of radius 1 degree around the relic. The metallicity was frozen to 0.2 solar, while the redshift was set to the redshift of NGC 4839, $z=0.0246$ \citep{Smith2000}. The factor that accounts for calibration errors was calculated with respect to the MOS2 spectrum. The RESP power indices and normalizations were coupled between spectra of the same detector, taking into account that the normalizations differ only by a constant factor that can be calculated from the output of the {\sc xmm-esas} routines.

\begin{figure*}
  \begin{center}
    \includegraphics[width=0.325\textwidth,clip=true,trim=0.2cm 0.5cm 1.3cm 0cm]{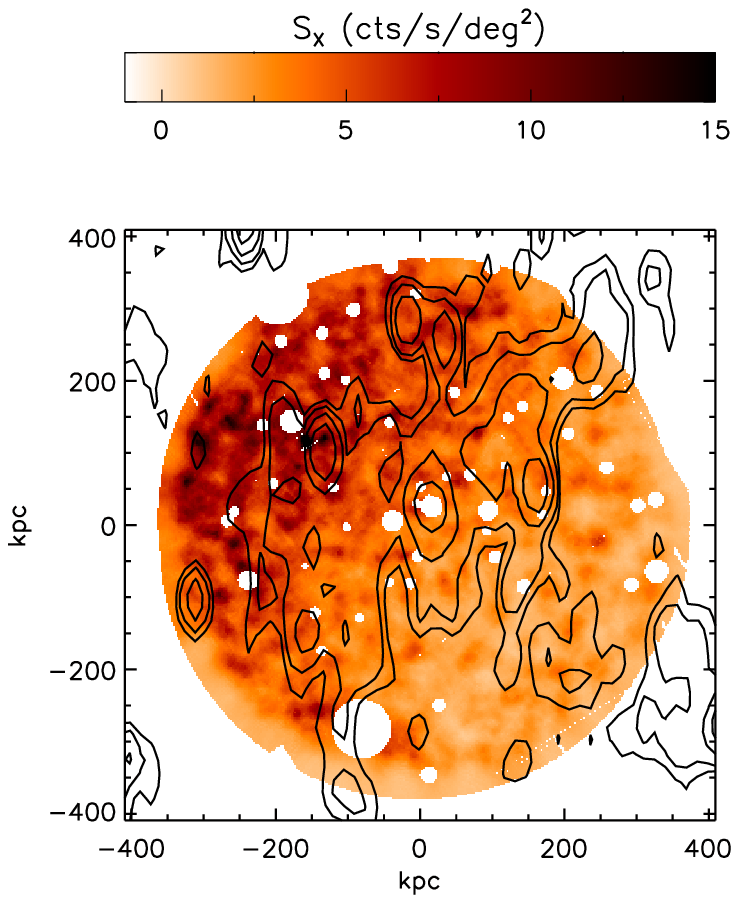}
    \includegraphics[width=0.325\textwidth,clip=true,trim=0.2cm 0.5cm 1.3cm 0cm]{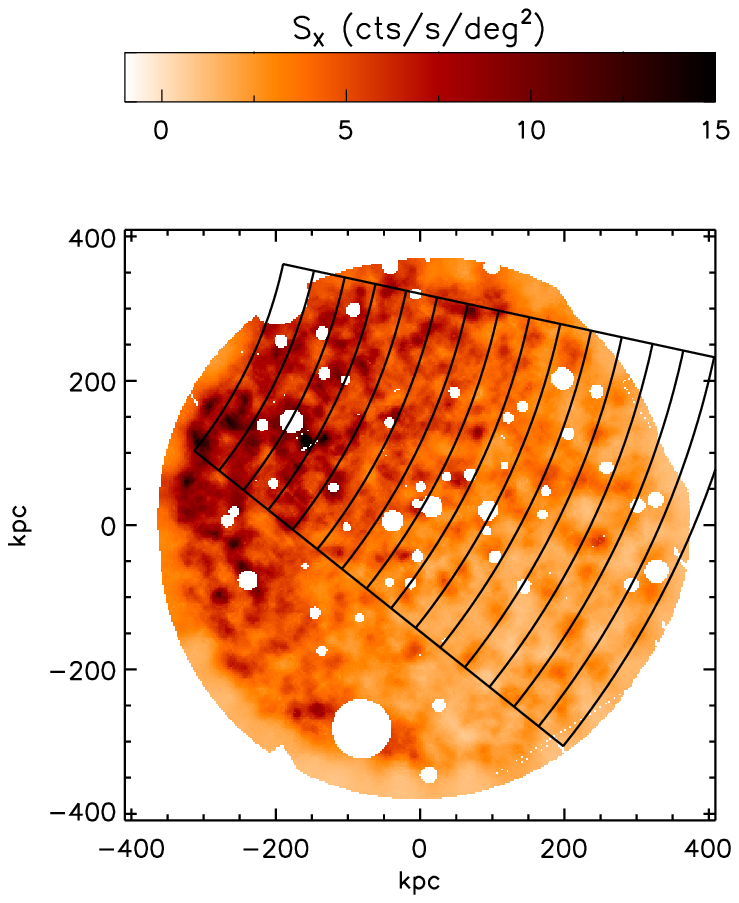}
    \includegraphics[width=0.325\textwidth,clip=true,trim=0.2cm 0.5cm 1.3cm 0cm]{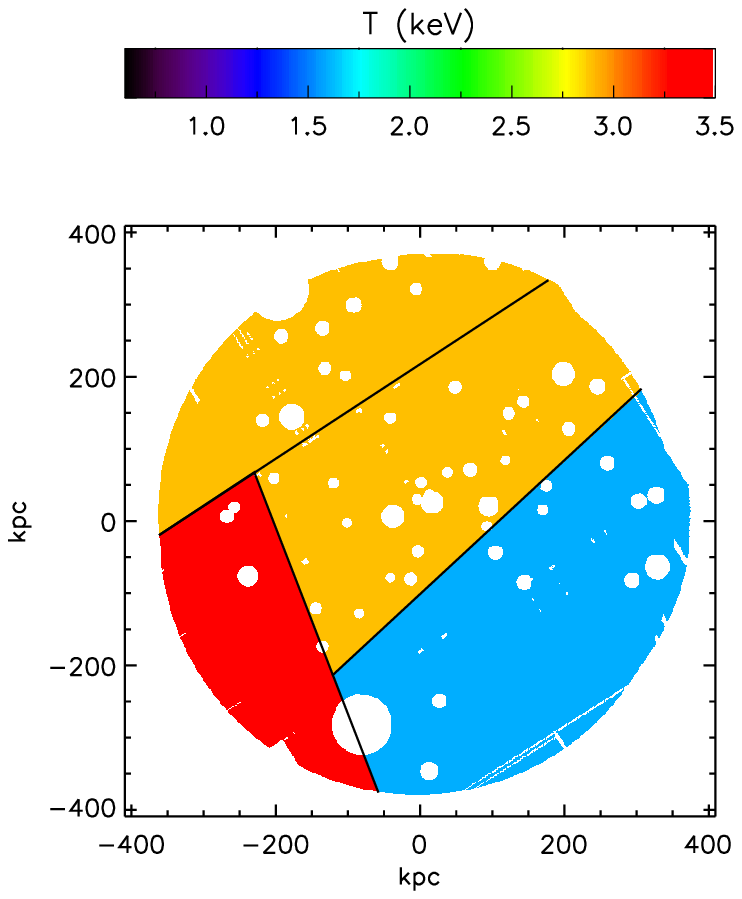}
  \end{center}
  \caption{\emph{Left:} Instrumental background-subtracted surface brightness map of the ICM near the Coma relic, binned by a factor of 2 and adaptively smoothed. The image corresponds to the energy band $0.4-4$ keV. Overlaid are WSRT 90-cm radio contours. The circular region has a radius of 12 arcmin. \emph{Centre:} The same X-ray image, now showing the partial annuli used for the surface brightness profile. \emph{Right:} Temperature map. Black lines separate the regions used for spectral analysis.}
\label{fig:xmm}
\end{figure*}

The count statistics of the \xmm\ observation only allow measuring the ICM temperature to within $\sim 10$ percent in about four regions. We chose them to cover a region on the relic, and regions in front and behind the relic. Because the surface brightness appears enhanced to the SE, we modelled the ICM in this region separately. The regions are shown in Figure \ref{fig:xmm}. The spectra were all grouped to a minimum of 30 counts per spectral channel, and fitted simultaneously in the same energy band as the sky background spectra. By fitting all the spectra simultaneously, the RESP parameters, the only ones linked between spectra, are better constrained. The best-fit spectral model for each region is summarized in Table \ref{tab:regions}. No SP component has been detected in the pn spectra ($\mathcal{R}$ for the pn data is consistent with 1.15 within the error bars). The best-fit RESP parameters for the MOS1 spectrum of the inner 12 arcmin around the optical axis are $\Gamma=1.1_{-0.12}^{+0.088}$ and $\mathcal{N}=3.7_{-0.88}^{+0.86} \times 10^{-2}$, with the normalization in default {\sc XSpec} units. The best-fit calibration factors ranged between $0.73$ and $0.96$. The overall fit has $\chi^2/{\rm d.o.f.}=1.00$.

\begin{figure*}
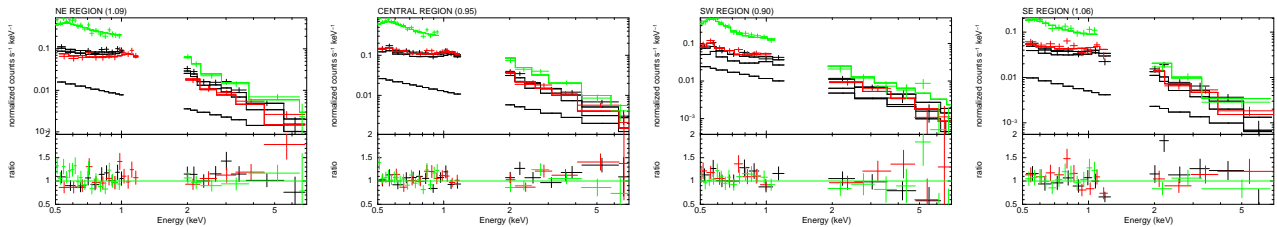

  \begin{center}
    \includegraphics[width=0.1665\textwidth,angle=270]{ne-sp.ps}
    \includegraphics[width=0.1665\textwidth,angle=270]{cntr-sp.ps}
    \includegraphics[width=0.1665\textwidth,angle=270]{sw-sp.ps}
    \includegraphics[width=0.1665\textwidth,angle=270]{se-sp.ps}
  \end{center}
  \caption{\emph{Top:} MOS1 (black), MOS2 (red), and pn (green) spectra of the NE, central, SW, and SE region, with overplotted best-fit models. Bottom panel shows the ratio of the data to the model. The power-law corresponds to the MOS1 RESP component, which was added to the model without being folded through the instrumental effective area. Reduced $\chi^2$ values for each fit are given in the plot title, in parenthesis.}
\label{fig:spectra}
\end{figure*}

Figure \ref{fig:spectra} shows the fitted spectra of the four regions. The NE and central regions have roughly equal temperature, $\approx 3$ keV, consistent with the previous temperatures calculated by \citet{FerettiNeumann2006} for regions NE of and overlapping the relic. However, we measure a strong temperature jump at the SW edge of the relic; the temperature here decreases to $\approx 1$ keV, which points towards a shock of Mach number $1.8_{-0.32}^{+0.22}$, derived using the Rankine-Hugoniot relations for $\gamma=5/3$.

\begin{table}
 \caption{Best-fit ICM models. Temperatures have units of keV, while normalizations ($\mathcal{N}$) are given in units of cm$^{-5}$\,arcmin$^{-2}$. For comparison, we also list the best-fit temperatures ($T^{\rm FN06}$) obtained by \citet{FerettiNeumann2006} for regions on the inner side of the relic (NE), and on the relic (centre); however, while their regions are included in ours, the exact region selection is different.}
\begin{center}
  \vspace{0.2cm}
 \begin{tabular}{lcccc}
   \multicolumn{4}{l}{\footnotesize{\textsc{SPECTRAL MODEL:}}} \\ 
   \multicolumn{4}{l}{\footnotesize{\textsc{calib $\times$ solid\_angle $\times$ ( phabs $\times$ apec )}}} \\ 
  \hline
   Region & $T$ & $\mathcal{N}$ & $T^{\rm FN06}$ \\
  \hline
    NE & $2.9\pm 0.21$ & $(1.7\pm 0.073) \times 10^{-5}$ &  $3.3_{-0.5}^{+0.7}$\\
    centre & $2.9\pm 0.21$ & $(9.3\pm 0.36) \times 10^{-6}$ & $3.3_{-0.4}^{+0.3}$ \\
    SW & $1.6_{-0.32}^{+0.20}$ & $(2.0\pm 0.30) \times 10^{-6}$ & -- \\
    SE & $3.4_{-0.32}^{+0.42}$ & $(1.3\pm 0.075) \times 10^{-5}$ & -- \\
    12 arcmin & $2.6_{-0.11}^{+0.12}$ & $(8.2\pm 0.21) \times 10^{-6}$ & -- \\
  \hline
 \end{tabular}
\end{center}
 \label{tab:regions}
\end{table}

We verified the robustness of our results by varying each background parameter within $1\sigma$ of its best-fit value, and refitting the ICM spectra with the new background model. The newly fitted temperatures are all consistent with the previous best-fit values, within their $1\sigma$ errors. Furthermore, using data from the Solar Wind Ion Composition Spectrometer on board of the Advanced Composition Explorer satellite, we also plotted the O$^{+7}$/O$^{+6}$ ion ratio as a function of time for one week before and after the date of the Coma and background observations, to see whether the data was affected by solar wind charge exchange (SWCX) \citep{Snowden2004}. The O$^{+7}$/O$^{+6}$ ratio during both of the \xmm\ observations was among the lowest in the two-week period, with values $\lesssim 0.1$.

\begin{figure}
  \begin{center}
    \includegraphics[width=\columnwidth, clip=true, trim=1cm 0.2cm 0cm 0.6cm]{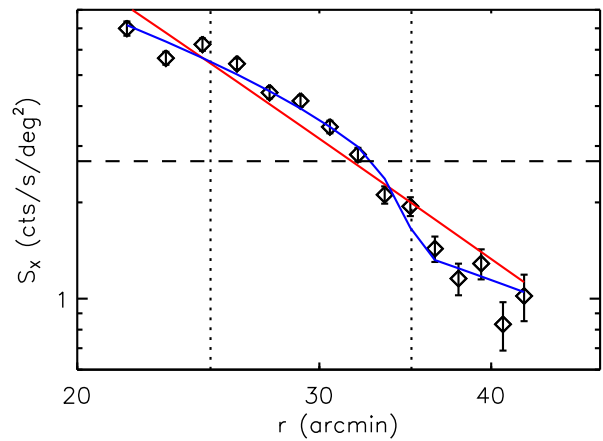}
  \end{center}
  \caption{Background-subtracted, vignetting-corrected surface brightness profile across the relic. The surface brightness was calculated in the energy band $0.4-4$ keV. Error bars show Poissonian errors. Dotted lines show the projected extent of the relic. The dashed horizontal line represents the background level (sky plus instrumental). Overplotted are best-fit models corresponding to an underlying power-law density profile (red; $\chi^2/{\rm d.o.f.}=7.9$), and two power-laws with equal indices and a jump fixed at a radius of 35 arcmin (blue; $\chi^2/{\rm d.o.f.}=3.0$).}
\label{fig:sx}
\end{figure}

\subsection{No surface brightness jump}

All shocks associated with radio relics were first identified based on discontinuities in the surface brightness (e.g., A3667, \citet{finoguenov10}; A754, \citet{Macario2011}). We extracted a surface brightness profile across the relic in partial annuli with an angular opening of 15 degrees and with a width of 15 arcmin, centred on NGC 4839. The profile is shown in Figure \ref{fig:sx}. It does not show any clear density discontinuity at the relic. Under the assumption of spherical symmetry, however, the surface brigthness profile is not consistent with a simple power-law density profile within the FOV (because the cluster's redshift is low, there is essentially no difference between a beta profile and a power-law profile on length scales equal to the \xmm's 30 arcmin FOV). Describing instead the density profile by two power-laws with a jump at a certain shock radius yielded improved, but still unsatisfactory fits. We tried fixing the density jump to the shock compression expected from the temperature-derived Mach number, but the fit remained poor. The best fit was obtained for power-laws with equal indices ($\Gamma=1.07$), and a jump fixed at $35$ arcmin; the shock compression ratio was approximately 1.2, but the fit had a $\chi^2/{\rm d.o.f.}$ of only $3$. 

In conclusion, the surface brightness profile does not appear consistent with $\mathcal{M}\approx 2$ or with an underlying beta-model density profile, at least not for the simple spherical models assumed here. 

\section{The nature of the shock}

SW of the Coma relic, we have identified a temperature jump that corresponds to a Mach number of $1.8_{-0.32}^{+0.22}$. However, no similar discontinuity is found in density, assuming a simple spherical-shock model. This result is not necessarily surprising though. At the position of the relic, the observed surface brightness profile is the integrated emissivity of both Coma's ICM, and the ICM of the infalling NGC 4839 group. This is a first deviation from our simplistic model, which will flatten the observed density profile. Furthermore, if the infall does not occur in the plane of the sky, then projection effects will smooth out the apparent density jump. Further smoothing can be caused by substructure along the line of sight. A similar effect has been seen in the merger A665, in which a shock region in front of the moving subcluster core has been identified in the temperature map, but is not associated with a surface brightness discontinuity \citep{MarkevitchVikhlinin2001,Govoni2004}.

The temperature jump measured at the relic corresponds to a shock of $\mathcal{M}\sim 2$. The nature of the shock/relic has been longly debated. \citet{Ensslin1998} suggested that the relic traces an accretion shock caused by ongoing gas accretion onto the cluster. \citet{FerettiNeumann2006} did not detect any evidence of a shock at the relic, and suggested that it might have been formed by turbulence.  Recently, \citet{BrownRudnick2011} proposed an alternative explanation: that the Coma relic traces an infall shock, the only infall shock claimed so far, caused by the infall of the NGC 4839 group onto Coma. Another possibility is that the shock is another more common outgoing merger shock.

It is now clear from optical and X-ray data that the NGC 4839 group is falling into Coma along a NE-SW-oriented cosmic filament, at a velocity of $\sim 1500$ ${\rm km\,s^{-1}}$ \citep{CollessDunn1996,Neumann2001}. The distance between the relic and NGC 4839 is approximately 900 kpc, which means that the group crossed the current position of the relic approximately 0.5 Gyr ago. If the shock is an outgoing merger shock, the temperatures on both sides of the relic imply a shock velocity of roughly 1200 ${\rm km\,s^{-1}}$. In this geometry, 0.5 Gyr ago the infalling group was at the current position of the relic, while the shock was in-between Coma and the NGC 4839 group, 0.6 Mpc ahead of the group. This cluster/group configuration would be consistent with the initial stage of a binary galaxy cluster merger predicted by numerical simulations, when a shock is formed at the interface between the two clusters. Numerical simulations also show that, following first core passage, outgoing merger shocks will be released into the ICM, outward of the two cluster cores; outgoing merger shocks are not produced behind the infalling subcluster core during its first infall onto the more massive cluster \citep[e.g.,][]{Takizawa2000,Ricker2001,vanWeeren2011b}. The NGC 4839 group is currently located approximately 1.5 Mpc away from Coma's centre, and it is observed well before first core passage. Therefore, the shock at the Coma relic, observed behind the infalling group, cannot be an outgoing merger shock triggered during a simple binary merger event between the group and Coma.

The low Mach number of the shock also excludes the possibility that we are observing an accretion shock projected at a radius smaller than Coma's virial radius. Accretion shocks are expected to have Mach numbers $\gtrsim 10$ \citep[e.g.,][]{Miniati2000}, while our \xmm\ data only reveals a shock of $\mathcal{M}\sim 2$.

The Coma relic has an integrated spectral index of 1.18 \citep{Thierbach2003}. Taking into account electron aging, the Mach number predicted by the injection spectral index is 3.5. The temperature discontinuity points towards a shock of Mach number $\sim 2$. However, colder gas from the cosmic filament must have been dragged into Coma's ICM in the wake of the infalling group, while the ICM of the group must have been ram-pressured stripped. Therefore, our assumption that the pre-shock and post-shock regions are best described by single-temperature thermal components is simplistic, and the actual strength of the shock is likely higher. Unfortunately, the low count statistics of our data do not allow testing two-temperature spectral models. Detailed spectral index and polarization mapping of the relic, as well as numerical simulations aimed at modelling the group's infall, are needed to further characterize the shock at the Coma relic. However, by finding evidence that the relic traces a shock front, we can confidently eliminate the scenario in which the relic was formed by turbulence. Of the remaining possibilities, the most likely one appears to be that the shock at the Coma relic is the only infall shock detected so far.

\section*{Acknowledgments}

We thank Silvano Molendi, Dominique Eckert, and Steven Snowden for their comments regarding the X-ray background analysis, Luigina Ferreti for kindly provinding the WSRT radio contours, and Annalisa Bonafede, Franco Vazza, and Elke Roediger for helpful suggestions. MB and FV acknowledge support from the Deutsche Forschungsgemeinschaft through grant FOR1254.

\bibliographystyle{mn2e}
\bibliography{bibliography}

\label{lastpage}

\end{document}